\documentclass[preprint]{aastex62}
\usepackage{booktabs}
\usepackage{makecell}
\usepackage{graphicx}
\usepackage{subcaption}  
\usepackage{amsmath}

\usepackage{xcolor}

\graphicspath{{./}{figures/}}

\shorttitle{3I/ATLAS encounter}
\shortauthors{Guo et al.}
\usepackage{subcaption} % 保留你想用的 subcaption（最后加载）

\makeatletter
\expandafter\let\csname longtable*\endcsname\relax
\expandafter\let\csname endlongtable*\endcsname\relax
\makeatother
\begin{document}

\title{Search for Past Stellar Encounters and the Origin of 3I/ATLAS}

\email{ffeng@sjtu.edu.cn (FF)}

\author{Yiyang Guo}
\affil{Department of Astronomy, School of Physics and Astronomy and Shanghai Key Laboratory for Particle Physics and Cosmology,
Shanghai Jiao Tong University, Shanghai 200240, People’s Republic of China}

\author{Luyao Zhang}
\affil{Department of Physics and Astronomy,
University of Leicester,
Leicester, LE1 7RH, UK}

\author[0000-0001-6039-0555]{Fabo Feng}
\affiliation{State Key Laboratory of Dark Matter Physics, Tsung-Dao Lee Institute \& School of Physics and Astronomy, 
Shanghai Jiao Tong University,
Shanghai 201210, China}
\affil{Department of Astronomy, School of Physics and Astronomy and Shanghai Key Laboratory for Particle Physics and Cosmology,
Shanghai Jiao Tong University, Shanghai 200240, People’s Republic of China}

\author[0000-0001-5017-7021]{Zhao-Yu Li}
\affil{Department of Astronomy, School of Physics and Astronomy and Shanghai Key Laboratory for Particle Physics and Cosmology,
Shanghai Jiao Tong University, Shanghai 200240, People’s Republic of China}
\affiliation{State Key Laboratory of Dark Matter Physics, Tsung-Dao Lee Institute \& School of Physics and Astronomy, 
Shanghai Jiao Tong University,
Shanghai 201210, China}

\author{Anton Pomazan}
\affil{Shanghai Astronomical Observatory, Chinese Academy of Sciences, Shanghai 200030, China}

\author{Xiaohu Yang}
\affil{Department of Astronomy, School of Physics and Astronomy and Shanghai Key Laboratory for Particle Physics and Cosmology,
Shanghai Jiao Tong University, Shanghai 200240, People’s Republic of China}
\affiliation{State Key Laboratory of Dark Matter Physics, Tsung-Dao Lee Institute \& School of Physics and Astronomy, 
Shanghai Jiao Tong University,
Shanghai 201210, China}

% \collaboration{(AAS Journals Data Scientists collaboration)}



\begin{abstract}
3I/ATLAS, the third discovered interstellar object, has a heliocentric speed of 58 $\mathrm{km\,s^{-1}}$ and exhibits cometary activity. To constrain the origin of 3I/ATLAS and its past dynamical evolution, we propagate the orbits of 3I/ATLAS and nearby stars to search for stellar encounters. Integrating orbits in the Galactic potential and propagating the astrometric and radial-velocity uncertainties of 30 million \textit{Gaia} stars, we identify 25 encounters with median encounter distances less than 1 pc. However, because the encounter speeds between 3I/ATLAS and each encounter exceed 20 $\mathrm{km\, s^{-1}}$, none is a plausible host under common ejection mechanisms. We infer stellar masses for most stars and quantify the gravitational perturbations exerted by each individual star or each binary system on 3I/ATLAS. The strongest gravitational scattering perturber is a wide M-dwarf binary. Among all past encounters, the binary's barycenter and 3I/ATLAS reach the small encounter distance of $0.242^{+0.089}_{-0.084}$~pc and the encounter speed of $28.39^{+0.67}_{-0.67}$ $\mathrm{km\,s^{-1}}$, 1.64 Myr ago. We further demonstrate that the cumulative influence of the stellar encounters on both the speed and direction of 3I/ATLAS is weak. Based on the present kinematics of 3I/ATLAS to assess its origin, we find that a thin-disk origin is strongly favored, because the thin disk both exhibits a velocity distribution closely matching that of 3I/ATLAS and provides the dominant local number density of stars.

\end{abstract}



\section{Introduction} \label{sec:intro}

Every interstellar object passing through the Solar System offers a valuable window for studying other star systems \citep{marvceta2023synthetic,hopkins2023galactic}. The first interstellar object 1I/'Oumuamua was discovered in 2017 and was confirmed to originate from outside the Solar System due to its hyperbolic trajectory. 1I/'Oumuamua has exhibited asteroidal features like a small, elongated shape and no detectable outgassing \citep{meech2017brief,vazan2020aspect,zhou2022observable}. To date, there is no
consensus on the mechanism that produced its shape or on its origin \citep{ye20171i,feng2018oumuamua,zheng2025configuration}. 
Mamajek~\citeyearpar{mamajek2017kinematics} and Gaidos et al.~\citeyearpar{gaidos2017origin} establish the kinematic baseline, showing that 1I/'Oumuamua’s velocity is unusually close to the local standard of rest (LSR). Other researchers have selected stars with well-measured positions and velocities and have traced the past million‑year orbits of 1I/'Oumuamua and of those stars to find encounter candidates \citep{dybczynski2018investigating}. 

The second interstellar visitor 2I/Borisov, characterized by a diffuse appearance and a homogeneous composition, was discovered in 2019 \citep{guzik2020initial}. The second interstellar object differs dramatically from the first in terms of cometary activity, shape and incoming direction \citep{bodewits2020carbon,prodan2024pre}. The difference between 1I/'Oumuamua and 2I/Borisov also hints that interstellar objects have diverse characteristics and origins \citep{pena2024oort,mikryukov2024rendez}.  \cite{hallatt2020dynamics} suggest that 2I/Borisov is older than 1I/'Oumuamua due to the higher speed and use a disk-heating model to find several encounters, although at high relative speeds. \cite{bailer2020search} also discuss the possibility of interstellar objects ejected from binary or planetary systems, and for old interstellar objects, it is more difficult to find their parent stars.

After a six-year gap, the third interstellar object 3I/ATLAS was discovered by the ATLAS survey in July 2025 \citep{seligman2025discovery}. Vera C. Rubin Observatory captured 3I/ATLAS in June 2025,  prior to its confirmation \citep{chandler2025nsf}.  Since its discovery, numerous observations have described 3I/ATLAS's features \citep{seligman2025discovery,bolin2025interstellar}. It is the brightest and largest interstellar comet observed so far \citep{bolin2025interstellar,de2025assessing}. Given its brightness and proximity, promptly constraining its origin will benefit subsequent months of observation and analysis. 3I/ATLAS is significantly faster than other interstellar objects with a heliocentric velocity of $ \sim 58\ \mathrm{km\,s^{-1}}$. The kinematics of 3I/ATLAS, together with the metallicities of possible host stars, imply that it may have formed at low metallicities and in the early age of the Galaxy \citep{taylor2025kinematic}. One study suggests the velocity of 3I/ATLAS determines a thick-disk origin with an age of $\sim$9.6 Gyr and the other study including 3I/ATLAS's chemistry and dynamics predicts an age over 7.6 Gyr and a high water mass fraction \citep{kakharov2025galactic,hopkins2025different}.
In general, 3I/ATLAS's unique velocity and size would indicate a different origin from the two previously discovered interstellar objects.

In this work, we aim to
find 3I/ATLAS's stellar encounters and understand its origin. We select a catalog of 30 million stars from \textit{Gaia} DR3 with good astrometric and radial velocity measurements and propagate the orbits of these stars and of 3I/ATLAS in the galactocentric coordinate frame. With a close-approach threshold of 1 pc, we ultimately identify 25 stars and analyze the dynamical effect of those close encounters. From the present kinematics of 3I/ATLAS, its three-dimensional velocity is more consistent with the velocity dispersion of thin-disk stars than with that of the thick-disk population.

The paper is structured as follows. We introduce the data used to propagate orbits in Section \ref{sec:data}. Then we explain how we calculate encounters in several stages and list the encounter results in Section \ref{sec:integ}. To analyze whether those encounters are the hosts of 3I/ATLAS, we introduce their interaction with 3I/ATLAS in Section \ref{Sec:ind}. We further model the Gyr-scale velocity and direction diffusion from encounters and determine 3I/ATLAS's origin is likely in the thin disk in Section \ref{sec:cum}. Finally, we discuss and conclude in Section \ref{sec:results}.

\section{Data Source and Selection} \label{sec:data}

We use data from \textit{Gaia} Data Release 3 (DR3; \citealt{gaia23}), which provides astrometric and radial velocity measurements for 33,812,183 sources. Each source includes five-parameter astrometry—Right Ascension (R.A.; $\alpha$), Declination (Decl.; $\delta$), parallax ($\varpi$), proper motion in R.A. ($\mu_{\alpha}$), and proper motion in Decl. ($\mu_{\delta}$)—as well as radial velocity ($v_r$).

After applying the parallax zero-point correction \citep{2021A&A...649A...4L,2024A&A...691A..81D} and accounting for systematic biases in proper motions \citep{cantat2021characterizing}, we select a subset of $\sim$30,000,000 stars with high-quality measurements as candidate encounter stars. These sources have valid radial velocity data and satisfy a fractional parallax error criterion of $0 < \sigma_{\varpi} / \varpi < 0.2$ \citep{2015PASP..127..994B, Zhang_2025}.

Accurate orbital integration requires well-constrained initial conditions in six-dimensional phase space. As \textit{Gaia} DR3 provides astrometric and kinematic measurements in the International Celestial Reference System, which is centered on the Solar System barycenter, we convert these observables to galactocentric Cartesian phase-space coordinates, adopting the Galactic parameters and initial conditions of the Sun outlined in
\citet{zhou2023circular}. 

To align the initial epoch of 3I/ATLAS with the \textit{Gaia} DR3 reference epoch, we query JPL Horizons\footnote{\url{https://ssd.jpl.nasa.gov/horizons/app.html}} at the epoch J2016.0 and retrieve the corresponding heliocentric positions and velocities. With continued observations, the state vector and its associated uncertainties will be gradually refined. In this work, the heliocentric state vector at the epoch J2016.0 is $(X,\,Y,\,Z)=\bigl(47.0454\pm0.0157,\,-111.6559\pm0.0181,\,5.0709\pm0.0010\bigr)$ au and $(V_X,\,V_Y,\,V_Z)=\bigl(-23.2454\pm0.0079,\,53.1991\pm0.0089,\,-2.3727\pm0.0005\bigr)$ km $\mathrm{s^{-1}}$.

\section{Orbital Integration} \label{sec:integ}

Given the vast size of the \textit{Gaia} DR3 catalog, the majority of stars are unlikely to experience close encounters with 3I/ATLAS. To reduce computational complexity, we first apply a linear approximation to estimate encounter times and distances assuming straight-line motion as a pre-filter. This allows us to select a subset of stars with small predicted encounter distances, which we then use as input for more computationally intensive nonlinear orbital integration. Finally, we estimate uncertainties in the orbital parameters using a Monte Carlo method.

\subsection{Preliminary Selection of Encounters}

Assuming constant relative velocity $\mathbf{u}$ between a star and 3I/ATLAS, the linear encounter time is given by the formalism of \citet{2015A&A...575A..35B}:
\begin{equation}
t^{\mathrm{lin}}_{\mathrm{enc}} = -\frac{\mathbf{r}_0 \cdot \mathbf{u}}{\mathbf{u} \cdot \mathbf{u}}~,
\end{equation}
where $\mathbf{r}_0$ is the initial relative position vector. The corresponding encounter distance is:
\begin{equation}
d^{\mathrm{lin}}_{\mathrm{enc}} = \left| \mathbf{r}_0 + \mathbf{u} \, t^{\mathrm{lin}}_{\mathrm{enc}} \right|~.
\end{equation}

We first select encounter candidates with nominal distance $d^{\mathrm{lin}}_{\mathrm{enc}} < 50$~pc, yielding $\sim$350,000 candidates. As shown by \citet{2015A&A...575A..35B}, adopting a sufficiently large threshold on linear encounter distance $d^{\mathrm{lin}}_{\mathrm{enc}}$ does not miss targets that would qualify as genuine close encounters in subsequent nonlinear propagation.

 We perform orbital integration using the \texttt{galpy} package \citep{2015ApJS..216...29B}, over a $\pm100$~Myr time span with a timestep of 2000 yr. After computing discrete orbital trajectories, we apply cubic spline interpolation to extract the periastron parameters. We test orbital propagation with time steps of 2000, 1000, and 100 yr and find negligible differences in encounter results.

In nonlinear orbit calculations, the Galactic gravitational potential plays a dominant role in shaping stellar orbits over long timescales. To ensure accurate orbital integration, we adopt the Milky Way mass model of \citet{zhou2023circular}, which is constrained by rotation curves derived from APOGEE and LAMOST data \citep{majewski2017apache,zhao2012lamost}.
As most stellar encounters produce minimal dynamical perturbations, we assume a smooth, static Galactic potential and neglect local gravitational effects such as spiral density waves and interstellar structures \citep{2021AstL...47..180B}. At this stage, a total of $\sim$3,400 stars with the nominal encounter distance $d^{\mathrm{nom}}_{\mathrm{enc}} < 10$~pc are retained as encounter candidates. 

\subsection{Selection of Encounters through the Monte Carlo approach}

Following the Monte Carlo (MC) method for astrometric sampling described in \citet{Zhang_2025}, we generate 1,000 star clones for each encounter candidate with distance $d^{\mathrm{}}_{\mathrm{enc}}$ less than 10 pc and 200 3I/ATLAS clones by sampling their state-vector uncertainties. In the previous study, errors in the Galactic potential model and in the Sun's position and kinematics contribute little to the encounter‑trajectory uncertainty compared with the star's own parameters \citep{feng2019probabilistic}.   We then apply the same  \texttt{galpy} orbital integration procedure.  Because the stellar and 3I/ATLAS samplings are statistically independent, every star clone can be paired with every 3I/ATLAS clone, yielding 200,000 MC realizations. Finally, we identify 25 stars with median encounter distance $d^{\mathrm{med}}_{\mathrm{enc}}$ less than 1 pc as close encounters. To describe the
 significance of perturbation of each encounter, we use a  perturbation proxy $g$ which is introduced by \cite{feng2015finding} based on numerical 
simulations of the perturbation from stellar encounters. For an encounter with mass of $M_{\star}$, encounter distance of $d^{\mathrm{}}_{\mathrm{enc}}$ and encounter speed $v^{\mathrm{}}_{\mathrm{enc}}$ at the close-approach moment, $g$ is:
 \begin{equation}\label{Eq:g}
     g=\dfrac{M_{\star}}{{d^{\mathrm{}}_{\mathrm{enc}}}^{2}\,v^{\mathrm{}}_{\mathrm{enc}}}
.
 \end{equation}

\begin{figure}[htbp]
	\centering
	
	\includegraphics[width=0.9\linewidth]{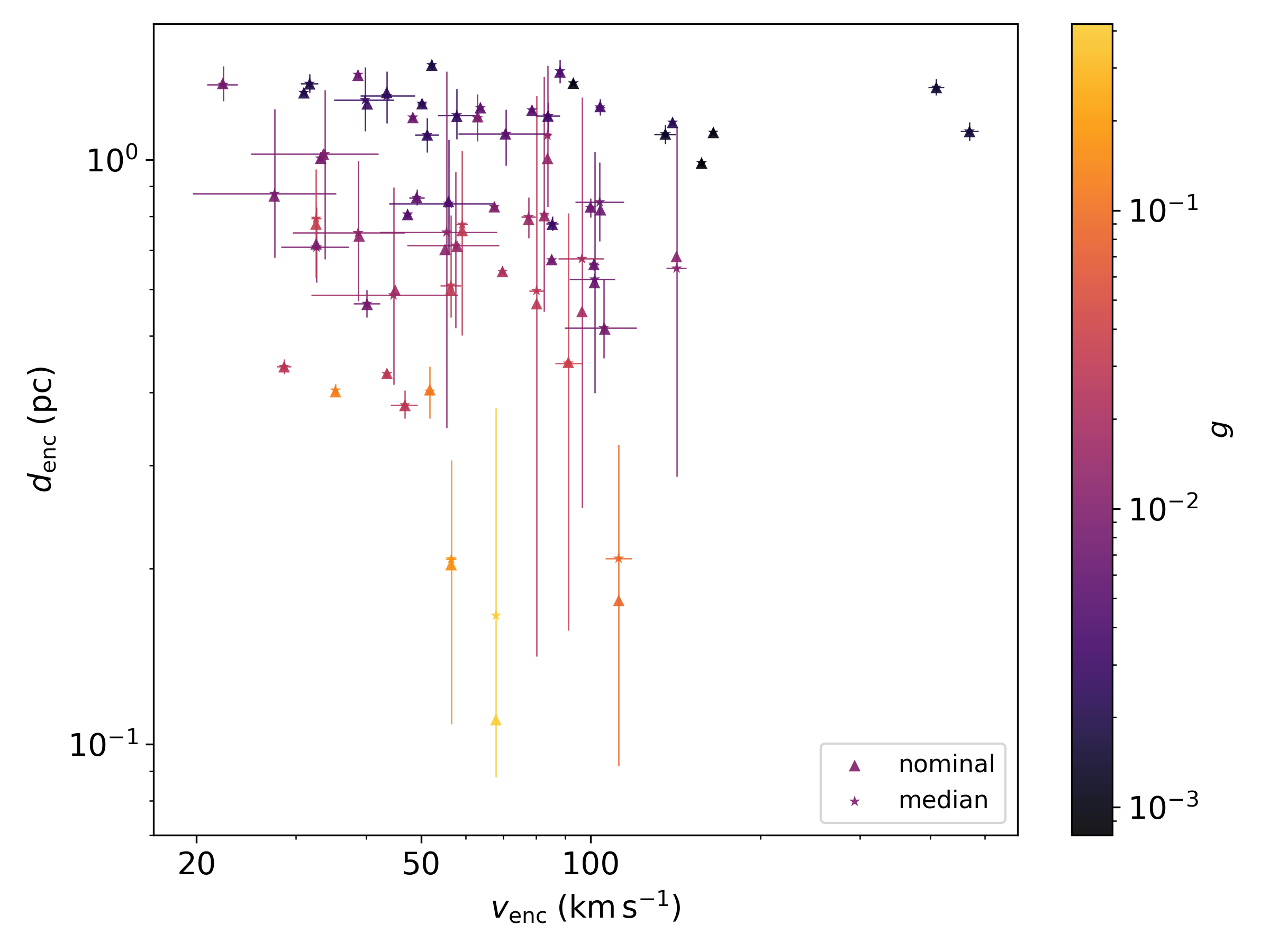}

  \caption{Distributions of nominal and median encounter distances $d^{\mathrm{}}_{\mathrm{enc}}$ and encounter speeds $v^{\mathrm{}}_{\mathrm{enc}}$ for encounters with each 95\% quantile of $d^{\mathrm{}}_{\mathrm{enc}}$ below 1.5 pc. The left panel represents the nominal values of $d^{\mathrm{nom}}_{\mathrm{enc}}$ and
$v^{\mathrm{nom}}_{\mathrm{enc}}$. 
The right panel represents median values of $d^{\mathrm{med}}_{\mathrm{enc}}$ and
$v^{\mathrm{med}}_{\mathrm{enc}}$.  Error bars indicate 5\% and 95\% quantiles of $d^{\mathrm{}}_{\mathrm{enc}}$ and $v^{\mathrm{}}_{\mathrm{enc}}$ in MC results.
The color bar is based on Equation \ref{Eq:g} using the median of $g^{\mathrm{}}$.}
  \label{fig:1pc}         
\end{figure}

Figure \ref{fig:1pc} shows the distributions of both nominal and median distances $d^{\mathrm{nom}}_{\mathrm{enc}}$ and velocities $v^{\mathrm{med}}_{\mathrm{enc}}$.  
It shows that the nominal encounter distance $d^{\mathrm{nom}}_{\mathrm{enc}}$ and the median encounter distance $d^{\mathrm{med}}_{\mathrm{enc}}$ generally differ only modestly. Only in some cases the nominal value is noticeably biased toward the 5th-percentile encounter distance $d^{5\%}_{\mathrm{enc}}$, and thus is not representative of the optimal estimate. So we prefer to use the median encounter distance $d^{\mathrm{med}}_{\mathrm{enc}}$ less than 1 pc as a threshold to select 25 stars. The only criterion simultaneously guarantees a small true close-approach distance while keeping the MC distance dispersion relatively low. Some encounter masses $m_{\mathrm{enc}}$ are available in \textit{Gaia} dataset via Final Luminosity Age Mass Estimator \citep{gaia23}, otherwise we interpolate  from a mass-luminosity relation given by \cite{pecaut2013intrinsic}.  Among the 25 stars, some are in binary systems, so when propagating their orbits, we use the barycentric initial conditions of the binary system.
 The 20 single  encounters with the median encounter distance $d^{\mathrm{med}}_{\mathrm{enc}}$ less than 1 pc are listed in Table \ref{table:table1}.  

 We flag binary components if any of the following hold: (i) cross-match to the \textit{Gaia} eDR3 wide-binary catalog \citep{el2021million}; (ii) entry in the Washington Double Star (WDS) catalog \citep{mason20012001}; (iii) Gaia DR3 Non-Single-Star (NSS) solution \citep{gosset2025gaia}; (iv) Renormalised Unit Weight Error (RUWE) $>$  1.4 \citep{gaia23}. We then obtain three reliable wide‑binary pairs (five of the 25 stars) from the \textit{Gaia} eDR3 wide-binary catalog, listed in Table \ref{Table2}.  To obtain the barycentric initial conditions, we apply the same MC sampling to get each component star galactocentric position $\vec{x}_{1}$, $\vec{x}_{2}$ and velocities $\vec{v}_{1}$, $\vec{v}_{2}$. The system’s barycenter is then calculated by the mass-weighted sums:
 \begin{equation}
\vec{x}_{\mathrm{bary}}
  = \frac{M_{1}\,\vec{x}_{1} + M_{2}\,\vec{x}_{2}}{M_{1} + M_{2}},
\qquad
\vec{v}_{\mathrm{bary}}
  = \frac{M_{1}\,\vec{v}_{1} + M_{2}\,\vec{v}_{2}}{M_{1} + M_{2}} .
 \end{equation}

\begin{deluxetable*}{llccc ccc ccc ccc}  % ← 12 列：Order + g + Gaia + 9 组列
\tablecaption{The 20 single  encounters ($d_{\mathrm{enc}}<1$ pc) of 3I/ATLAS and their median, nominal, 5\% and 95\% quantiles of $t_{\mathrm{enc}}$, $d_{\mathrm{enc}}$ and $v_{\mathrm{enc}}$. \label{table:table1}}

\tabletypesize{\footnotesize}

\tablehead{
% ────────── 第 1 行：列标题 ──────────
\colhead{\rule{0pt}{2.6ex}Order} &
\colhead{\textit{Gaia} DR3 ID} &
% \colhead{$g^{\mathrm{med}}$} & \colhead{$g^{\mathrm{5\%}}$} & \colhead{$g^{\mathrm{95\%}}$} &
\colhead{$t_{\mathrm{enc}}^{\mathrm{med}}$} & \colhead{$t_{\mathrm{enc}}^{\mathrm{nom}}$} &\colhead{$t_{\mathrm{enc}}^{\mathrm{5\%}}$} & \colhead{$t_{\mathrm{enc}}^{\mathrm{95\%}}$} &
\colhead{$d_{\mathrm{enc}}^{\mathrm{med}}$} & \colhead{$d_{\mathrm{enc}}^{\mathrm{nom}}$} &\colhead{$d_{\mathrm{enc}}^{\mathrm{5\%}}$} & \colhead{$d_{\mathrm{enc}}^{\mathrm{95\%}}$} & 
\colhead{$v_{\mathrm{enc}}^{\mathrm{med}}$} & \colhead{$v_{\mathrm{enc}}^{\mathrm{nom}}$} &\colhead{$v_{\mathrm{enc}}^{\mathrm{5\%}}$} & \colhead{$v_{\mathrm{enc}}^{\mathrm{95\%}}$}
\\
% ────────── 第 2 行：单位 ──────────
\colhead{} & \colhead{} &
\colhead{Myr} &\colhead{Myr} &\colhead{Myr} &\colhead{Myr} &
\colhead{pc} & \colhead{pc} &  \colhead{pc} &  \colhead{pc} & 
\colhead{km\,s$^{-1}$} & \colhead{km\,s$^{-1}$} &\colhead{km\,s$^{-1}$} &\colhead{km\,s$^{-1}$}  
}

\startdata
1 & 5944464849163504128 & -1.62 & -1.62 & -1.64 & -1.61 & 0.166 & 0.110 & 0.088 & 0.377 & 67.89 & 67.88 & 67.42 & 68.36 \\
2 & 5909461003111201792 & -0.77 & -0.77 & -0.82 & -0.73 & 0.208 & 0.176 & 0.092 & 0.325 & 112.05 & 112.09 & 106.11 & 118.15 \\
3 & 6792436799477051904\textsuperscript{*} & -0.20 & -0.20 & -0.22 & -0.19 & 0.380 & 0.380 & 0.361 & 0.403 & 46.82 & 46.82 & 44.18 & 49.34 \\
4 & 6570039342736534784 & -0.81 & -0.81 & -0.82 & -0.81 & 0.401 & 0.404 & 0.361 & 0.442 & 51.82 & 51.83 & 51.65 & 51.99 \\
5 & 6863591389529611264 & -0.71 & -0.71 & -0.72 & -0.71 & 0.403 & 0.401 & 0.394 & 0.414 & 35.26 & 35.27 & 35.03 & 35.47 \\
6 & 6779821003058453120 & -0.33 & -0.33 & -0.34 & -0.32 & 0.442 & 0.442 & 0.430 & 0.456 & 28.58 & 28.59 & 27.74 & 29.40 \\
7 & 6698224978148515456 & -2.72 & -2.72 & -2.86 & -2.57 & 0.448 & 0.450 & 0.157 & 0.810 & 91.22 & 91.23 & 86.45 & 96.40 \\
8 & 6855915149098312064 & -0.58 & -0.58 & -0.68 & -0.51 & 0.516 & 0.513 & 0.458 & 0.625 & 105.42 & 105.87 & 89.93 & 120.55 \\
9 & 6580068022653080704 & -3.34 & -3.34 & -3.45 & -3.24 & 0.597 & 0.567 & 0.141 & 1.287 & 80.12 & 80.17 & 77.73 & 82.55 \\
10 & 6660744825780671744\textsuperscript{*} & -3.04 & -3.05 & -3.18 & -2.91 & 0.661 & 0.250 & 0.141 & 1.715 & 83.06 & 83.02 & 79.84 & 86.23 \\
11 & 6766620129016117888 & -4.30 & -4.29 & -4.72 & -3.92 & 0.677 & 0.550 & 0.254 & 1.279 & 96.45 & 96.47 & 87.59 & 105.43 \\
12 & 6849401917092877440 & -4.30 & -4.28 & -5.54 & -3.52 & 0.715 & 0.670 & 0.373 & 1.742 & 43.72 & 43.87 & 33.83 & 53.31 \\
13 & 6722801914901880064 & -1.97 & -1.98 & -2.59 & -1.60 & 0.751 & 0.702 & 0.348 & 1.416 & 55.49 & 55.15 & 42.24 & 68.17 \\
14 & 6767564510724876288 & -2.08 & -2.08 & -2.14 & -2.02 & 0.798 & 0.791 & 0.735 & 0.861 & 77.54 & 77.58 & 75.32 & 79.86 \\
15 & 4382898024013196928 & -3.16 & -3.16 & -3.22 & -3.10 & 0.805 & 0.802 & 0.550 & 1.387 & 82.69 & 82.71 & 81.29 & 84.13 \\
16 & 4189858726030433792 & -0.61 & -0.61 & -0.61 & -0.60 & 0.830 & 0.831 & 0.817 & 0.843 & 67.43 & 67.45 & 66.50 & 68.40 \\
17 & 6710764530307734656 & -1.44 & -1.44 & -1.59 & -1.30 & 0.846 & 0.821 & 0.726 & 0.990 & 103.71 & 104.05 & 93.76 & 114.48 \\
18 & 6397423335799106048 & -3.84 & -3.84 & -3.99 & -3.69 & 0.862 & 0.558 & 0.316 & 2.192 & 85.75 & 85.86 & 82.59 & 89.13 \\
19 & 6741607618172465152 & -4.51 & -4.51 & -4.60 & -4.42 & 0.925 & 0.851 & 0.514 & 1.528 & 86.24 & 86.26 & 84.83 & 87.67 \\
20 & 5254061535106490112 & -0.03 & -0.03 & -0.03 & -0.03 & 0.987 & 0.987 & 0.973 & 1.000 & 157.20 & 157.19 & 155.26 & 159.41 \\
\enddata
\tablecomments{
The encounters are sorted in increasing order of $d_{\mathrm{enc}}^{\mathrm{med}}$. The symbol \textsuperscript{*} denotes stars very likely to be binary components. }

\end{deluxetable*}

\begin{figure}[htbp]
	\centering
	\begin{subfigure}{0.487\linewidth}
		\centering
		\includegraphics[width=0.9\linewidth]{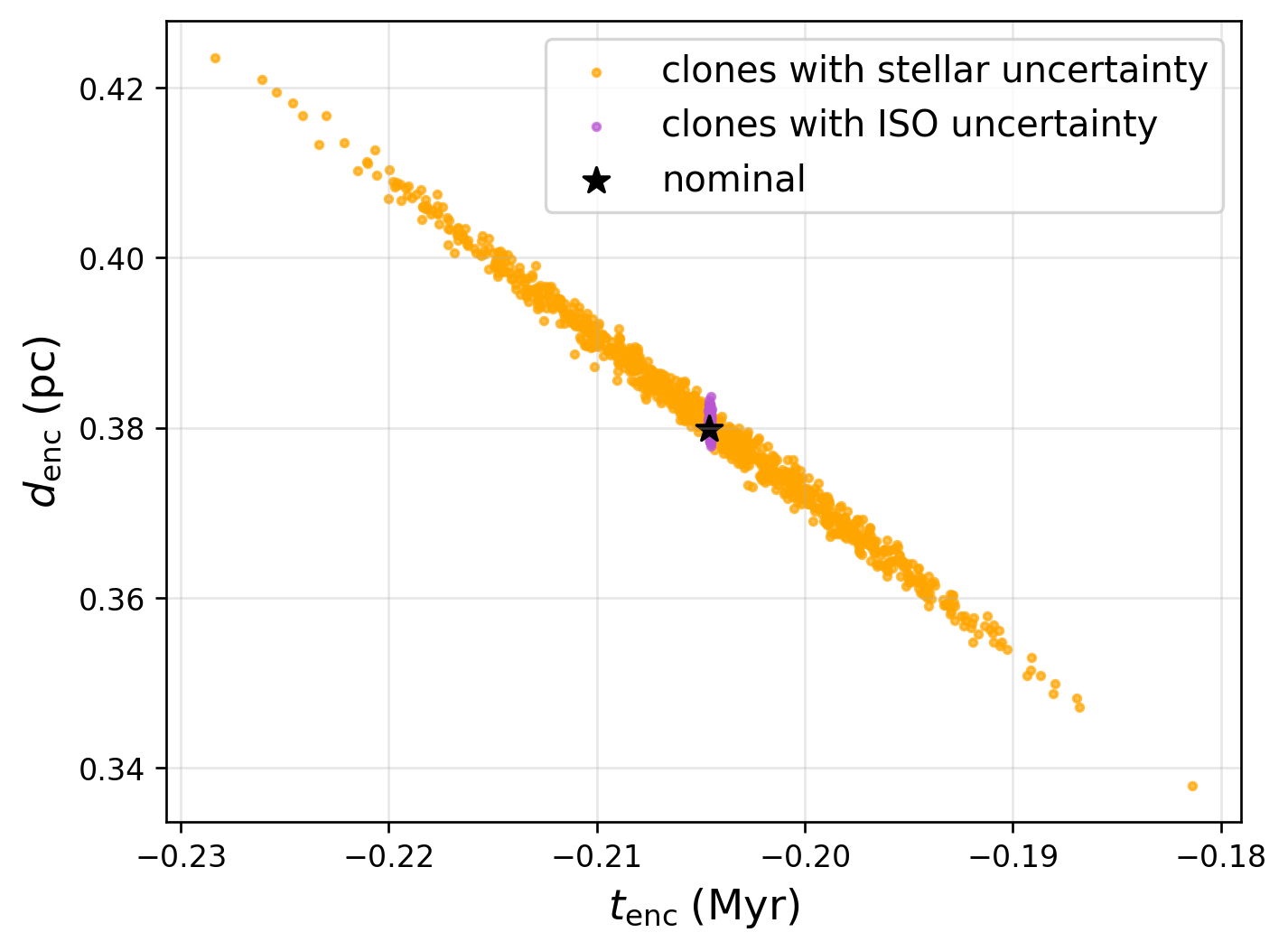}

	\end{subfigure}
	\centering
	\begin{subfigure}{0.487\linewidth}
		\centering
		\includegraphics[width=0.9\linewidth]{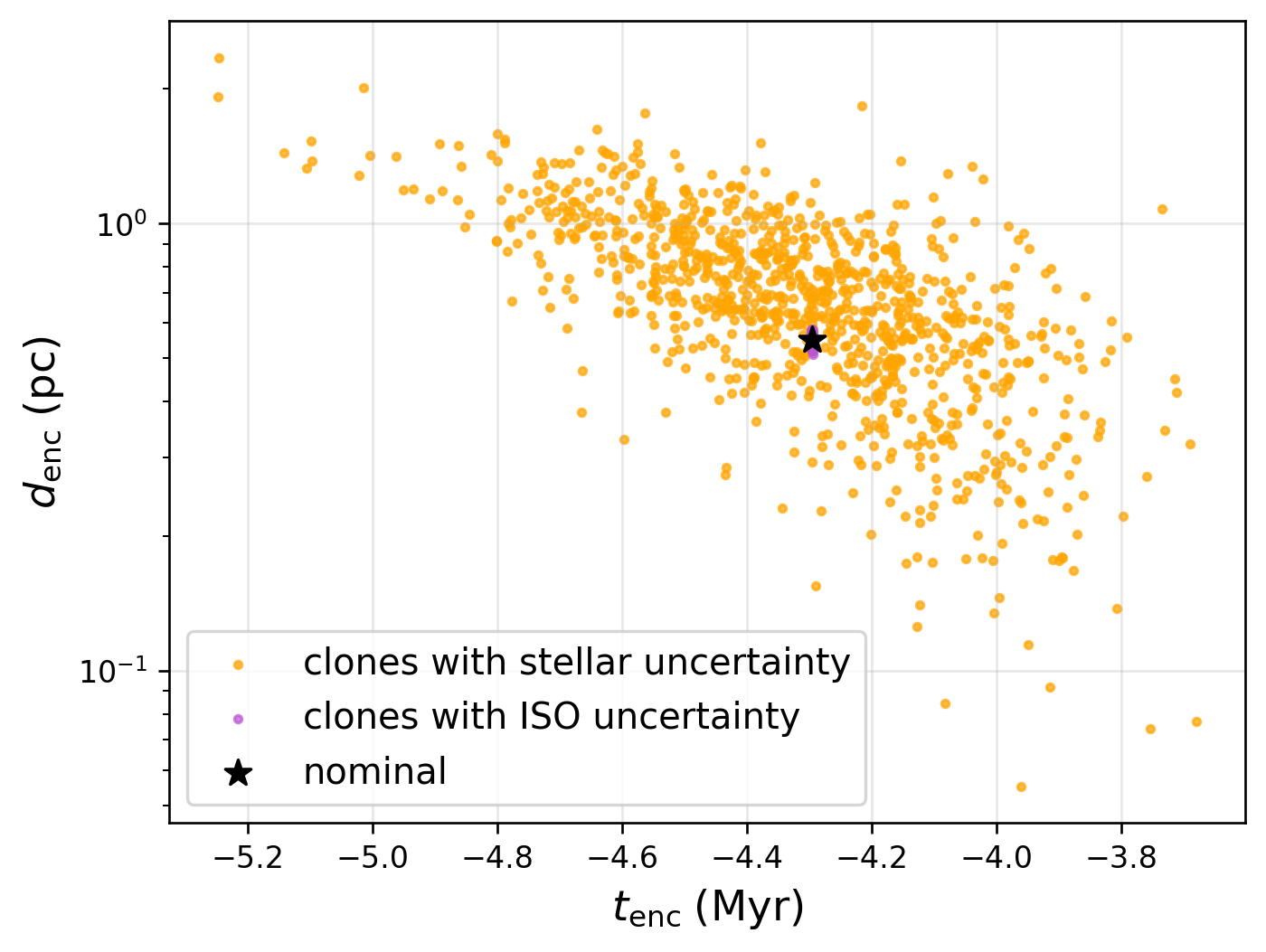}

	\end{subfigure}
  \caption{Comparison of two encounter MC results. In both panels, the orange dots represent one 3I/ATLAS nominal sample and 200 star MC samples pairwise results while the purple dots represent 1000 3I/ATLAS MC samples and 1 star nominal sample pairwise results. }
  \label{fig2:twostars}  
\end{figure}

Each encounter uncertainty is evaluated with 200,000 MC samples. In this ensemble, the dispersion of $d_{\mathrm{enc}}^{\mathrm{}}$ exhibits  a strong correlation with the absolute value of 
$t_{\mathrm{enc}}^{\mathrm{}}$. For stars with $|t^{\mathrm{}}_{\mathrm{enc}}| > 10$~Myr, the dispersion becomes much larger because of the present stellar parameter uncertainties. Besides, even in the early discovery phase of 3I/ATLAS, the uncertainties in the position and velocity of 3I/ATLAS are far smaller than uncertainties associated with the stellar parameters. Figure \ref{fig2:twostars} illustrates how the uncertainties from 3I/ATLAS and from the stellar parameters affect the error of encounters. From this figure, the dispersion of the purple points is significantly smaller than that of the orange points. This indicates that in long time span of orbital propagation, the calculation error bar is further dominated by the uncertainties in the star astrometry and radial velocity measurements rather than by the uncertainty of the pre-entry velocity and position of 3I/ATLAS itself.  

\section{Individual Encounters}\label{Sec:ind}
We summarize several historical encounters with the ten smallest median encounter distances $d^{\mathrm{med}}_{\mathrm{enc}}$ to further analyze the origin of 3I/ATLAS and the influence of encounter history. We treat all binary barycenters as single perturbers, since their projected separations (tens to thousands of au) are negligible compared with the parsec-scale values of $d_{\mathrm{enc}}$.

For the pair \textit{Gaia} DR3 1197546390909694720 (G 137-55) and \textit{Gaia} DR3 1197546562708387584 (G 137-54), we carry out a more detailed investigation.
Its projected separation is about 1358 au, placing it among ultra-wide binaries, and we analyze how the barycenter of this pair gravitationally scatters the orbit of 3I/ATLAS.
G 137-55 and G 137-54 form a binary system composed of two M-dwarf stars whose relative velocity is $0.197\ \mathrm{km\,s^{-1}}$ \citep{mason20012001,lepine2011all}. We provide an analytic expression for the deflection angle in a single scattering. We treat the barycenter and 3I/ATLAS as an ideal two-body problem, working in the
barycentric inertial frame.
We introduce the notation
of the gravitational constant $G$ and the stellar mass $M_\star$, the relative position $\mathbf r$ and the pre-encounter velocity $\mathbf v_{\mathrm{enc}}$ of 3I/ATLAS and the  barycenter . The impact parameter is:
\begin{equation}
    b = \frac{|\mathbf r \times \mathbf v_{\mathrm{enc}}|}{|\mathbf v_{\mathrm{enc}}|}.
\end{equation}
The  deflection angle is:
\begin{equation}
  \delta 
    = 2\arctan\!\Bigl(\frac{GM_\star}
    {b\,|\mathbf v_{\mathrm{enc}}|^{2}}\Bigr).
  \label{eq:deflect}
\end{equation}

\begin{figure}[htbp]
	\centering
	\begin{subfigure}{0.487\linewidth}
		\centering
		\includegraphics[width=0.9\linewidth]{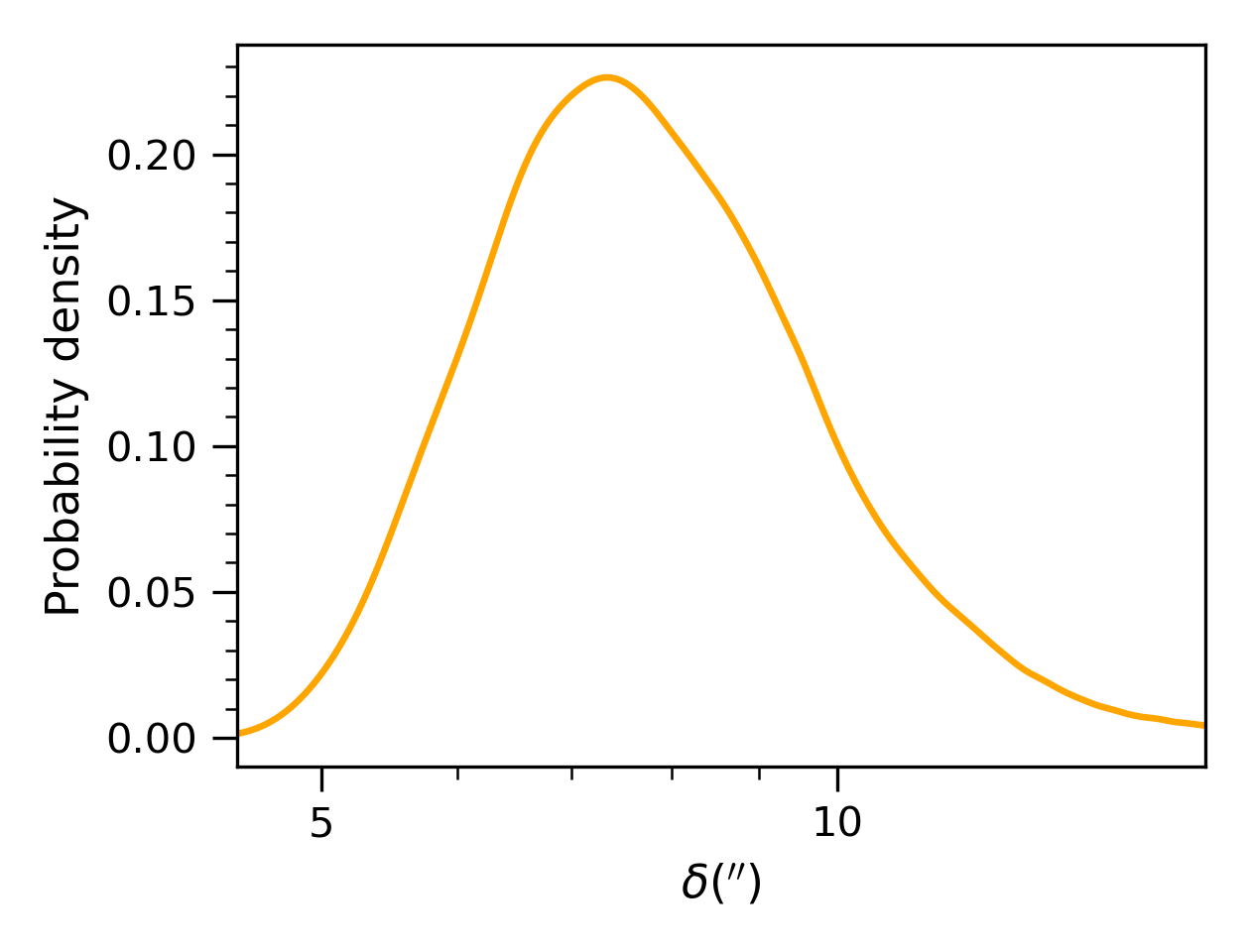}

	\end{subfigure}
	\centering
	\begin{subfigure}{0.487\linewidth}
		\centering
		\includegraphics[width=0.9\linewidth]{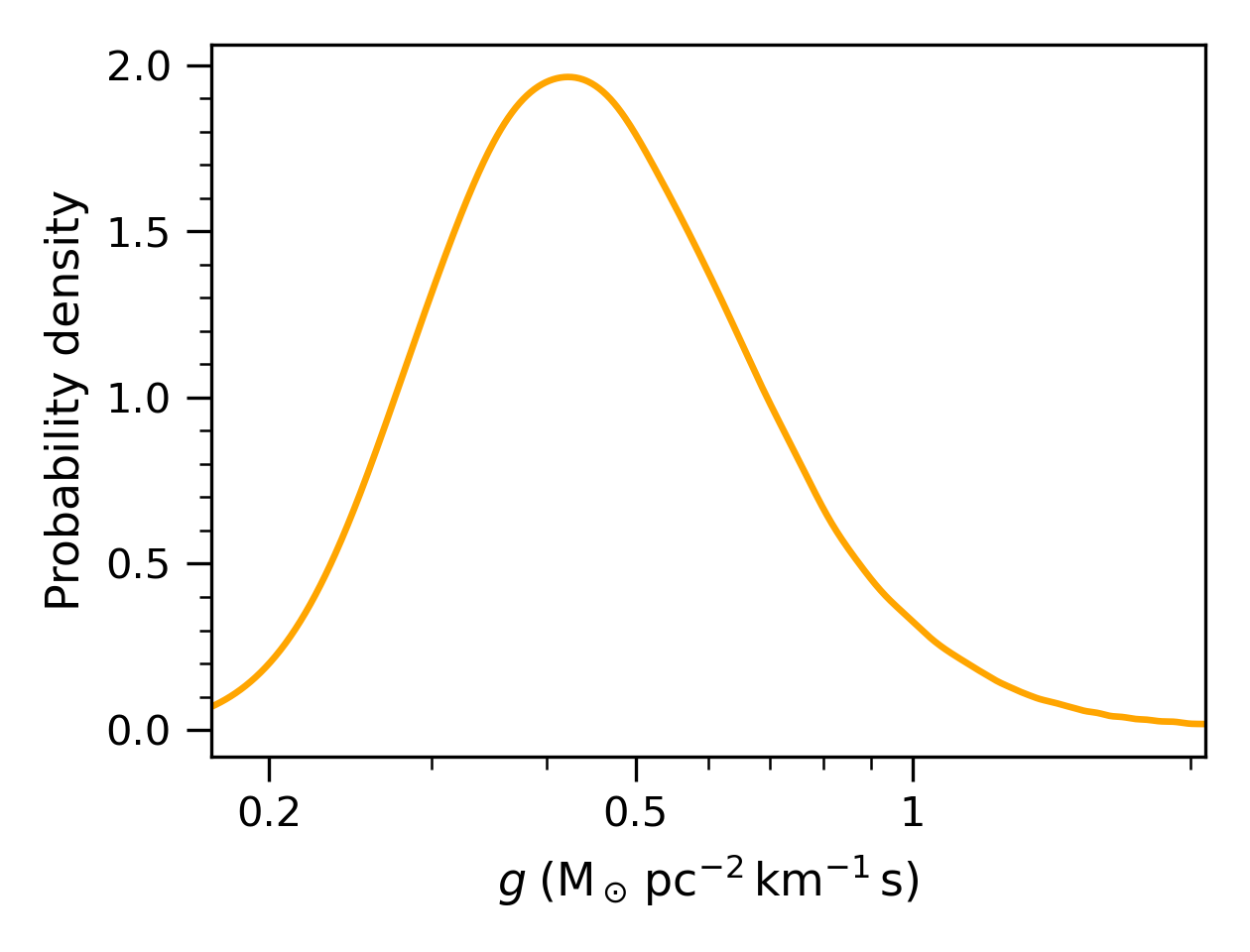}

	\end{subfigure}
  \caption{Distributions of deflection angle $\delta$ and $g$ smoothed with a Gaussian kernel.}
  \label{Fig:3}
\end{figure}

In the preceding set of 200,000 MC barycenter and 3I/ATLAS encounters, we present two indicators in Figure \ref{Fig:3}. This binary is the strongest perturber along the past 10 Myr segment of 3I/ATLAS’s trajectory. However, existing binary star ejection mechanisms, while capable of producing interstellar objects, are unlikely to explain the encounter speed over 20 km\,s$^{-1}$ \citep{jackson2018ejection,bailer2020search}.

\textit{Gaia} DR3 6792436799477051904 (V* AT Mic B) is a nearby star, and together with
\textit{Gaia} DR3 6792436799475128960 (V* AT Mic A) forms a binary system.  V* AT Mic A and  V* AT Mic B are both M-dwarfs  with the $\sim30$\,au projected separation \citep{joy1974spectral,malkov2012dynamical}, belonging to the $20$-$25$\,Myr $\beta$\,Pictoris moving group at 9.8 pc \citep{messina2017beta}. Their rapid rotation, strong X-ray and ultraviolet flaring \citep{Kuznetsov_2023}, and an ALMA-detected compact cold-dust belt \citep{cronin2023alma} suggest an active, planetesimal-rich environment that could, in principle, eject comets via binary or planetary perturbations.  Nevertheless, the system’s space motion lies within a few km\,s$^{-1}$ from the LSR, far below 3I/ATLAS's 
speed with respect to the LSR, and dynamical studies suggest the comet originated from an old, thick-disk star rather than a $\sim$25 Myr young dwarf \citep{taylor2025kinematic,hopkins2025different}.

\textit{Gaia} DR3 5909461003111201792  and \textit{Gaia} DR3 6855915149098312064   both have large median encounter speeds $v_{\mathrm{enc}}^{med}$.  In \textit{Gaia}~DR3, their radial velocities are $72.66$ and $48.45\ \mathrm{km\,s^{-1}}$.  \textit{Gaia} DR3 5909461003111201792 has an effective temperature of $ 3524\ \mathrm{K}$, whereas  \textit{Gaia} DR3 6855915149098312064 lacks a reliable mass and effective temperature estimate.  From a mass-luminosity relation, these two objects are consistent with M-dwarf classifications \citep{pecaut2013intrinsic}.  Such low-mass stars are unlikely to generate the large ejection velocities required for 3I/ATLAS and the large relative velocities  actually make both gravitational and non-gravitational encounter effects vanishingly small.

\textit{Gaia} DR3 5944464849163504128, \textit{Gaia} DR3 6863591389529611264, \textit{Gaia} DR3 6570039342736534784 and \textit{Gaia} DR3 6779821003058453120 all have reliable spectral types in \textit{Simbad}\footnote{\url{https://simbad.u-strasbg.fr/simbad/}} and are classified as K0, K5Vk, G3V, and M5e, respectively \citep{nesterov1995henry,gray2006contributions,gaidos2014trumpeting}. The G-type star could plausibly host giant planets \citep{feng2019search}. 
Such massive planets enhance the efficiency of ejecting small bodies which strong three–body scattering and repeated close encounters allow transfer of orbital energy and angular momentum to planetesimals, raising them to unbound, hyperbolic trajectories \citep{hinkel2019recommendation}. 
Nevertheless, the comparatively high encounter speeds and the stellar masses of all four systems, taken together, make it unlikely that any of them served as the host star of 3I/ATLAS.
 Moreover, at $d_{\mathrm{enc}}$ of 0.1 pc, the likelihood that 3I/ATLAS would interact with the giant planets or Oort‐cloud region is negligibly small.

{\renewcommand{\arraystretch}{1.5} 
\setlength{\tabcolsep}{5pt}      

\begin{deluxetable*}{ll cc ccc ccc ccc ccc}
\tablecaption{Binary-barycenter encounters with at least one component approaches 3I/ATLAS with $d_{\mathrm{enc}}^{\mathrm{med}}<1\,$pc in the \textit{Gaia} eDR3 wide-binary catalog.\label{Table2}}
\tabletypesize{\scriptsize}      % 比 \tiny 更稳妥，换字体时不易破版
\tablewidth{0pt}                    % ★ 让表格按“自然宽度”，横线随内容走
\tablehead{
\colhead{\rule{0pt}{2.6ex}Order} &
\colhead{\textit{Gaia} DR3 ID} &
\colhead{Mass} & \colhead{Separation} &
\colhead{$g_{\mathrm{enc}}^{\mathrm{med}}$} & \colhead{$g_{\mathrm{enc}}^{5\%}$} & \colhead{$g_{\mathrm{enc}}^{95\%}$} &
\colhead{$t_{\mathrm{enc}}^{\mathrm{med}}$} & \colhead{$t_{\mathrm{enc}}^{5\%}$} & \colhead{$t_{\mathrm{enc}}^{95\%}$} &
\colhead{$d_{\mathrm{enc}}^{\mathrm{med}}$} & \colhead{$d_{\mathrm{enc}}^{5\%}$} & \colhead{$d_{\mathrm{enc}}^{95\%}$} &
\colhead{$v_{\mathrm{enc}}^{\mathrm{med}}$} & \colhead{$v_{\mathrm{enc}}^{5\%}$} & \colhead{$v_{\mathrm{enc}}^{95\%}$} \\
\colhead{} & \colhead{} & \colhead{$M_\odot$} & \colhead{au} &
\multicolumn{3}{c}{$M_\odot\,\mathrm{pc}^{-2}\,\mathrm{km}^{-1}\,\mathrm{s}$} &
\colhead{Myr} & \colhead{Myr} & \colhead{Myr} &
\colhead{pc} & \colhead{pc} & \colhead{pc} &
\colhead{km\,s$^{-1}$} & \colhead{km\,s$^{-1}$} & \colhead{km\,s$^{-1}$}
}
\startdata
1 & \shortstack[l]{1197546390909694720\\1197546562708387584}
  & {\raggedleft 0.890 $\pm$ 0.064\hspace{0.2em}} & 1358.3
  & 0.534 & 0.284 & 1.277
  & -1.64 & -1.68 & -1.60
  & 0.242 & 0.158 & 0.331
  & 28.39 & 27.72 & 29.06 \\
2 & \shortstack[l]{4591398521365845376\\4591395600788082816}
  & {\raggedleft 0.353 $\pm$ 0.025\hspace{0.2em}} & 1370.0
  & 0.043 & 0.039 & 0.048
  & -1.07 & -1.12 & -1.03
  & 0.567 & 0.534 & 0.601
  & 25.50 & 24.40 & 26.55 \\
3 & \shortstack[l]{2386898972054841088\\2386898937694609920}
  & {\raggedleft 0.859 $\pm$ 0.062\hspace{0.2em}} & 152.3
  & 0.047 & 0.043 & 0.050
  & -0.53 & -0.57 & -0.49
  & 0.684 & 0.635 & 0.737
  & 39.22 & 36.43 & 42.26 \\
\enddata
\end{deluxetable*}
}
\section{Cumulative Effects of Encounter Scattering} \label{sec:cum}

\begin{figure}[htbp]
	\centering
	
	\includegraphics[width=0.52\linewidth]{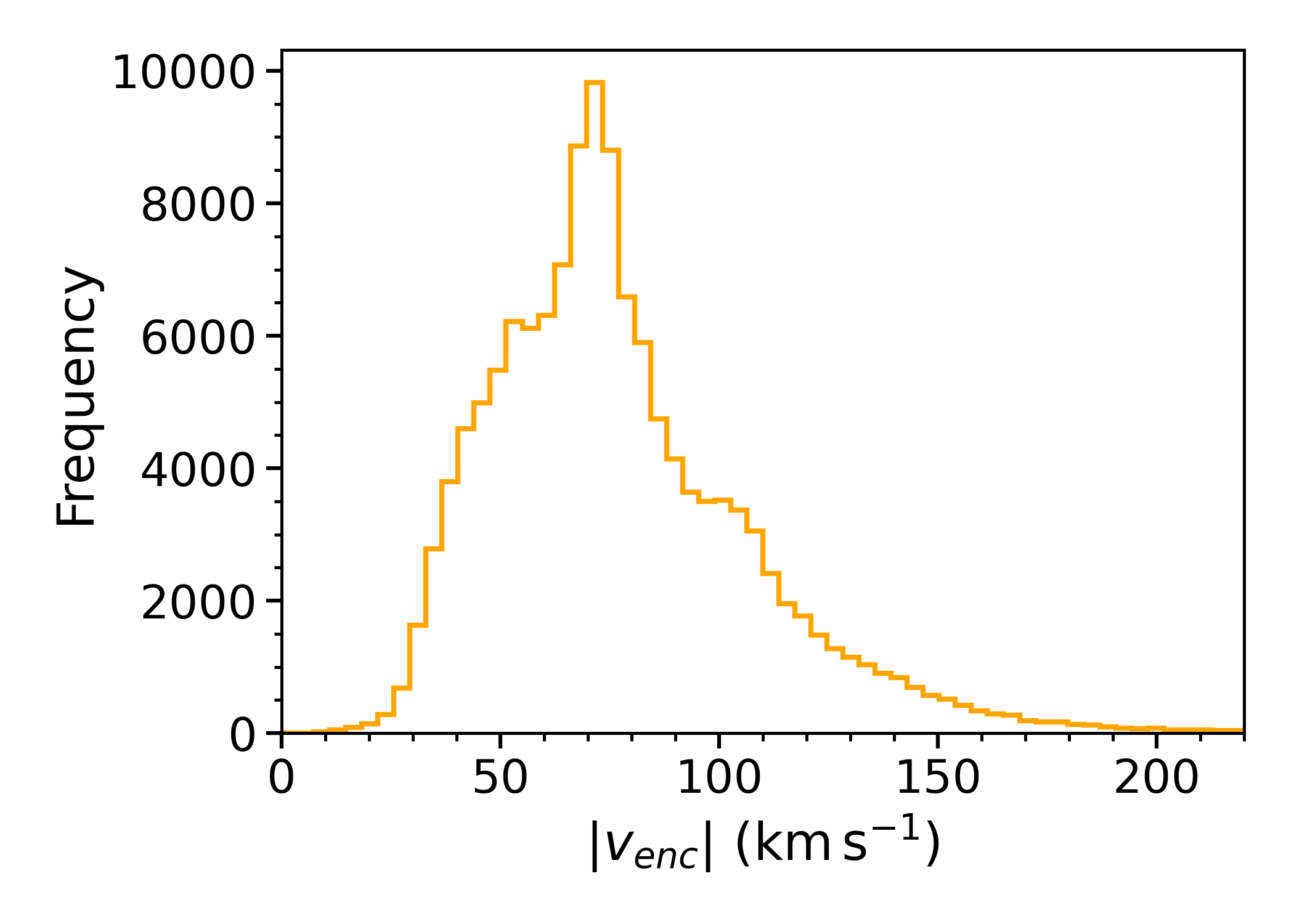}

  \caption{Histograms of relative speeds of 3I/ATLAS and  $\sim$134,000 nearby stars with radial velocities.}
  \label{Fig:4}
\end{figure}

We analyze each encounter’s gravitational slingshot to evaluate whether 3I/ATLAS originated near the Sun’s location, namely about 8 kpc from the Galactic
center and embedded in the thin disk, and subsequently was accelerated to its present high speed. We model 3I/ATLAS encounters by assuming that stars are distributed in a cylindrical tube of maximum radius $d_{\mathrm{max}}$ around the object's trajectory, and the encounter rate is uniformly distributed over time $T$. Following \cite{feng2014exploring}, the unnormalized probability density is:
\begin{equation}
    P(t_{\mathrm{enc}}, d_{\mathrm{enc}}, v_{\mathrm{enc}})= \alpha_{\rm ISO} \times \frac{2 d_{\mathrm{enc}}}{d_{max}^{2}} \times f_{v}( v_{\mathrm{enc}}),\label{eq:encp}
\end{equation}
where $\alpha_{\rm ISO}$ is 3I/ATLAS’s encounter rate and $f_{v}( v_{\mathrm{enc}})$ denotes the probability density of the encounter speed $v_{\mathrm{enc}}$ between the 3I/ATLAS and a star at the close approach moment. Figure \ref{Fig:4} shows the distribution of encounter speeds statistically derived from stars within 100 pc of 3I/ATLAS’s current position. We normalize Equation \ref{eq:encp}  by adopting the previously derived, completeness-corrected solar encounter rate $\alpha_{\odot}$ of \(21.8~\mathrm{Myr^{-1}}\) within \(1~\mathrm{pc}\) \citep{bailer2018new}. The solar encounter rate explicitly accounts for survey incompleteness by using a forward model of the stellar spatial, kinematic, and luminosity distributions together with an approximation to the \textit{Gaia} selection function.  The encounter rates of interstellar object $\alpha_{\mathrm{ISO}}$ and the Sun $\alpha_{\odot}$ are regarded as approximately  proportional to their speeds in the LSR-frame $v_{\mathrm{ISO}}$ and $v_{\odot}$:
\begin{equation}
    \frac{\alpha_{\mathrm{ISO}}}{\alpha_{\odot}}\approx\frac{v_{\mathrm{ISO}}}{v_{\odot}}.
\end{equation}

We then work in the LSR-frame to quantify how a single encounter
changes 3I/ATLAS’s velocity. Let $\mathbf v_\star$ denote the stellar velocity relative to the LSR, $\mathbf v_{\mathrm {enc}}$ and $\mathbf v_{\mathrm {enc}}'$ denote 3I/ATLAS’s pre-encounter velocity and post-encounter velocity relative to the star. The speed change $\Delta v_{}$ caused by a single encounter or a relatively close binary in the LSR-frame is:
\begin{equation}
    \Delta v=|\mathbf{v}_{\star}+\mathbf v_{\mathrm {enc}}'|-|\mathbf{v}_{\star}+\mathbf v_{\mathrm {enc}}|,
\end{equation}
where $\mathbf v_{\mathrm {enc}}'$ satisfies Equation \ref{eq:deflect}:
\begin{equation}
    cos\delta=\frac{\mathbf v_{\mathrm {enc}}' \cdot \mathbf v_{\mathrm {enc}}} {|\mathbf v_{\mathrm {enc}}'||\mathbf v_{\mathrm {enc}}|}, \,|\mathbf v_{\mathrm {enc}}'| = |\mathbf v_{\mathrm {enc}}|.
\end{equation}

\begin{figure}[htbp]
	\centering
	\begin{subfigure}{0.487\linewidth}
		\centering
		\includegraphics[width=0.9\linewidth]{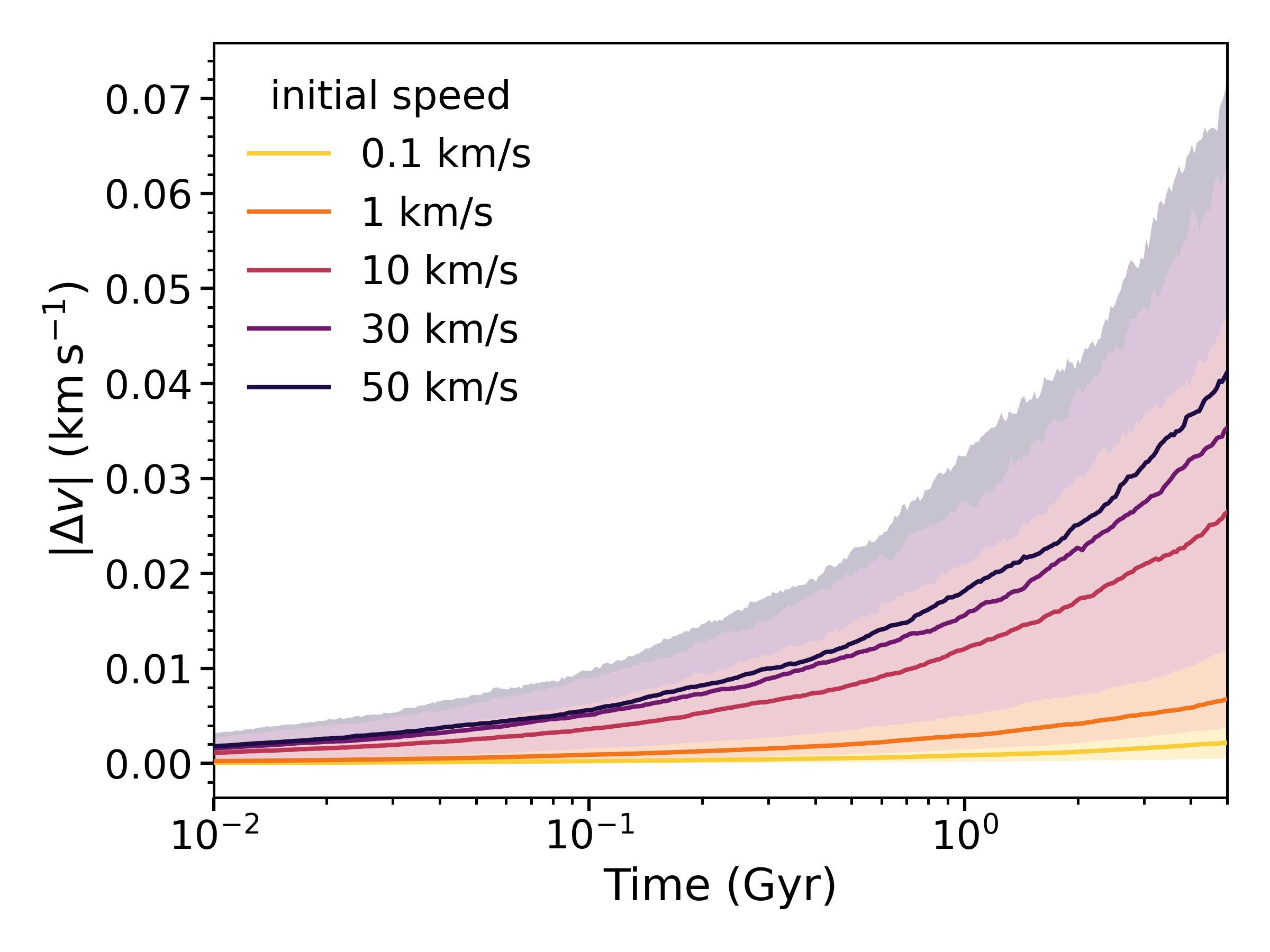}

	\end{subfigure}
	\centering
	\begin{subfigure}{0.487\linewidth}
		\centering
		\includegraphics[width=0.9\linewidth]{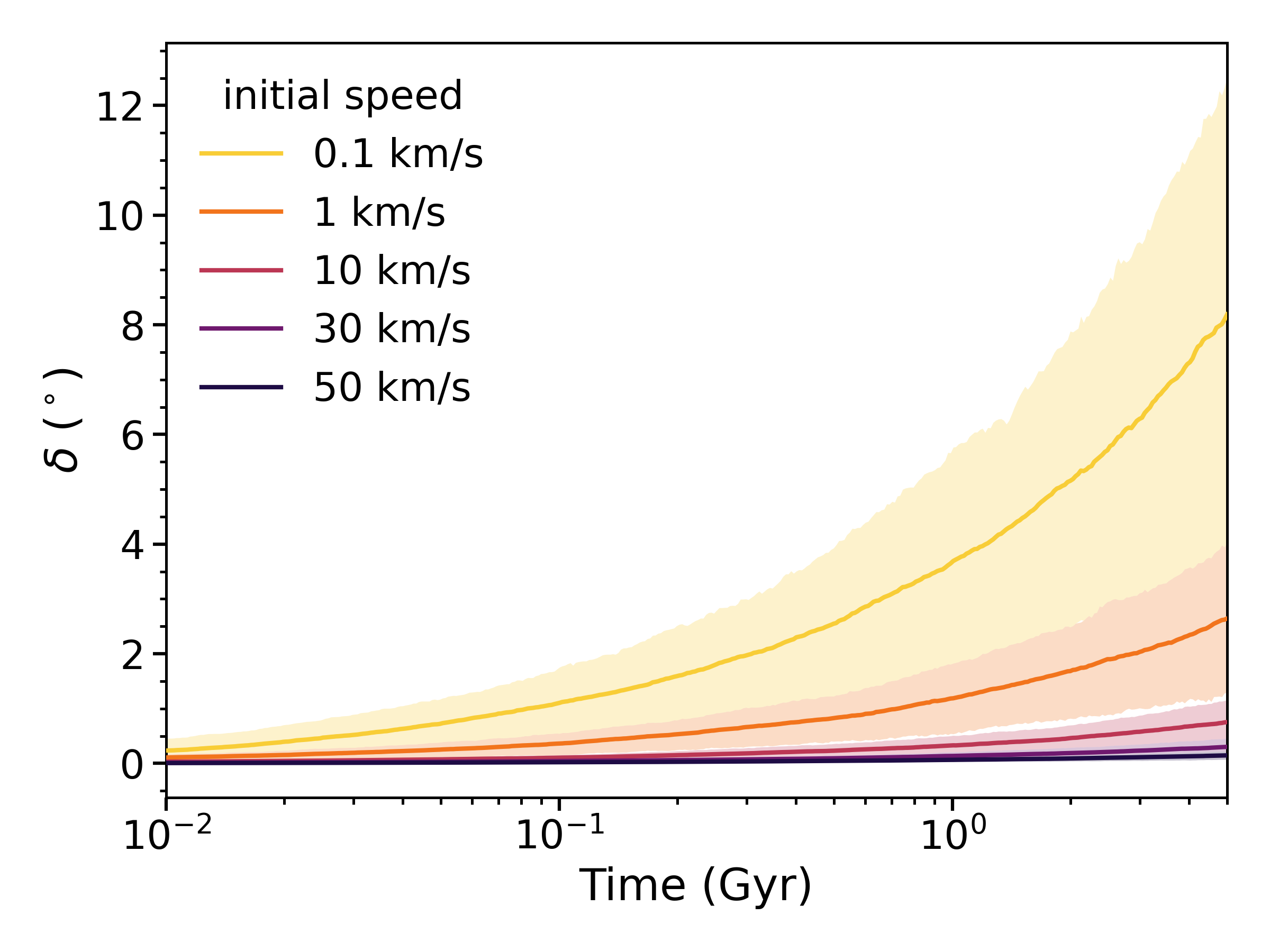}

	\end{subfigure}
  \caption{Cumulative speed change $\Delta v_{\rm }$ and deflection angle $\delta$.}
  \label{Fig:5}
\end{figure}

Accumulating all encounters with the maximum separation $d_{max} \le 1$ pc, we assess the gravitational scattering effects over 5 Gyr using 1000 MC realizations for several initial speeds. In these simulations, we assign a solar mass to each encountering star, although this likely overestimates the effect. 
Figure \ref{Fig:5} shows that 
the cumulative speed change  $|\Delta v|$, integrated over every encounter's gravitational scattering,  remains very small, while the
accumulated deflection angle $\delta$ shows a more appreciable change. 3I/ATLAS’s current speed in the LSR is around 58 $~\mathrm{km\,s^{-1}}$. 
Our model shows that the cumulative acceleration
from stellar slingshots over Gyr timescales is too small to boost an
object from canonical ejection speeds of order
\(\sim 10~\mathrm{km\,s^{-1}}\) up to \(\sim 50~\mathrm{km\,s^{-1}}\) \citep{pfalzner2021significant}.
At such high relative speeds, the scattering deflection is 
strongly suppressed, so both the speed and direction of 3I/ATLAS are only weakly
perturbed by random encounters. This
implies that 3I/ATLAS already possessed a large relative velocity in the LSR
and the heliocentric frame  at the time of ejection, rather than being
substantially accelerated afterward by stellar scattering.

To quantify the probability that 3I/ATLAS belongs to the thin or thick disk, we adopt the age–velocity–dispersion relation (AVR) measured from red-clump stars \citep{sun2025age}. The velocity dispersion $\sigma_{v}(\tau, R)$ is empirically modeled as a power law in age $\tau$ with a galactocentric radius $R$-dependent slope:
\begin{equation}
    \sigma_{v}(\tau, R)=\sigma_{v, 0}(R)[\tau+0.1]^{\beta_{v}(R)}.
\end{equation}
At the solar radius, we use the local kinematics of the thin disk and the thick disk from empirical fits in the standard galactocentric cylindrical coordinate $\{ R, \phi, z \}$: 
\begin{equation}
\begin{aligned}
 \langle v_{\phi}^{\mathrm{thin}}\rangle &= 224.82~\mathrm{km\,s^{-1}},\,
(\sigma_R^{\mathrm{thin}},\sigma_{\phi}^{\mathrm{thin}},\sigma_z^{\mathrm{thin}}) = (34.59,\,22.88,\,19.72)~\mathrm{km\,s^{-1}},\\
\langle v_{\phi}^{\mathrm{thick}}\rangle &= 176.94~\mathrm{km\,s^{-1}},\,
(\sigma_R^{\mathrm{thick}},\sigma_{\phi}^{\mathrm{thick}},\sigma_z^{\mathrm{thick}}) = (65.47,\,54.07,\,41.32)~\mathrm{km\,s^{-1}}.
\end{aligned}
\end{equation}
At the current galactocentric position and velocity of 3I/ATLAS, we evaluate likelihoods of the thin disk and the thick disk using the same functional form:
\begin{equation}
\mathcal{L}^{thin}(\mathbf{v})
=\frac{1}{(2\pi)^{3/2}\,\sigma_{R}^{thin}\,\sigma_{\phi}^{thin}\,\sigma_{z}^{thin}}
\exp\!\left[
-\frac{1}{2}\!\left(
\frac{v_R^{2}}{{\sigma_{R}^{\mathrm{thin}}}^{2}
}
+\frac{(v_{\phi}-v_{\phi}^{thin})^{2}}{{\sigma_{\phi}^{\mathrm{thin}}}^{2}}
+\frac{v_z^2}{{\sigma_{z}^{\mathrm{thin}}}^{2}}
\right)\right].
\end{equation}
The likelihood ratio $\mathcal{L}_{\rm thin}/\mathcal{L}_{\rm thick}$ is 4.43, which corresponds to about 81.6$\%$ probability for a thin-disk origin and 18.4$\%$ for a thick-disk origin without if we assume 3I/ATLAS's age $\tau$ is 5 Gyr and the same number density of stars in the thin or thick disk. Further including the local population fractions as priors,
$n_{\mathrm{thin}}$ of 0.8067 and $n_{\mathrm{thick}}$ of 0.1206 \citep{sun2025age}, the posterior probability of the thin-disk origin is:
\begin{equation}
    P_{thin}
= \frac{\ n_{thin}\,\mathcal{L}^{thin}(\mathbf{v})}
       { n_{thin}\,\mathcal{L}^{thin}(\mathbf{v})+ n_{thick}\,\mathcal{L}^{thick}(\mathbf{v})}.
\end{equation}
The thin-disk origin probability $P_{\mathrm{thin}}$ is $96.59\%$ and the thick-disk origin probability  $P_{\mathrm{thick}}$ is $3.41\%$. Assuming the age $\tau$ of 3I/ATLAS lies in the range \(3\) to \(15\) Gyr,
the posterior thin–disk probability \(P_{\mathrm{thin}}\) varies only weakly with age $\tau$,
from \(96.2\%\) to \(97.0\%\).
Hence, the conclusion is not sensitive to reasonable age variations.

Given the present kinematics of 3I/ATLAS, our results support an origin in the thin disk. If we restrict the evidence to the vertical component \(v_z\) alone, the thin-disk and thick-disk likelihoods are nearly equal corresponding to about 51.11$\%$ probability for a thin-disk origin and 48.89$\%$ for a thick-disk origin. Considering population fractions, the posterior origin probabilities of the thin disk $P_{thin}$ and the thick disk $P_{thick}$ are 87.43$\%$ and 12.57$\%$.
One study suggests a thick-disk origin of 3I/ATLAS based on maximum vertical excursion $z_{max}$ of  $0.480^{+0.023}_{-0.017}$ with a 68$\%$ confidence interval \citep{kakharov2025galactic}. 
However, this maximum vertical excursion is insufficient to unambiguously determine whether 3I/ATLAS belongs to the thin or thick-disk region \citep{yu2021mapping,chrobakova2022warp}. 
When the three-dimensional velocity is considered together with the relative numbers of stars in the thin and thick disk, a thin-disk origin becomes overwhelmingly more likely than a thick-disk origin.

\section{Discussion and Conclusion} \label{sec:results}
% \section{Discussion} \label{sec:discuss}

Our Gaia‑based search yields 25 close encounters, but their high relative speeds preclude a robust host‑star association. We divide the 25 stars into 20 single encounters and 3 binary encounter systems. An M-dwarf binary exerts the strongest gravitational scattering effect on 3I/ATLAS, with the barycentric median encounter distance of $0.242^{+0.089}_{-0.084}$ pc and  the median encounter speed of $28.39^{+0.67}_{-0.67}\,\mathrm{km\,s^{-1}}$. Both components of this binary show $g$ values that are an order of magnitude larger than those of any other candidate encounter, indicating that the system can exert the strongest gravitational perturbations on 3I/ATLAS. Although existing ejection mechanisms struggle to account for the very high ejection velocity of 3I/ATLAS, binaries provide additional dynamical pathways. 

To  clarify the magnitude of encounter perturbations to 3I/ATLAS, we use an analytic impulse model to calculate the deflection angle of G 137-55. Even though G 137-55 yields the largest significance of perturbation $g$,  the resulting deflection angle is only a few arcseconds. The current precision of star astrometry and radial velocity contributes much more uncertainty than scattering with the stars, especially when the encounter time $|t^{\mathrm{}}_{\mathrm{enc}}|$ is large. On Gyr timescales, the cumulative encounter-driven changes in both speed and direction are found to be very weak for 3I/ATLAS's current speed.
These considerations likewise support using nonlinear orbit integrations in a smooth galactic potential as a sufficiently accurate alternative to, in principle, full $N$-body modeling with stars and the galactic potential. At the million-body level, such $N$-body calculations are computationally prohibitive for our used time steps. On a 100 Myr timescale, impulses from close encounters and second-order star–star terms produce only negligible perturbations to the trajectory of 3I/ATLAS.

On a 100 Myr timescale, we adopt a static galactic potential. The Sun’s orbital period is around 220 Myr, and key time-dependent drivers evolve on longer or comparable scales. For example the vertical phase-mixing timescale associated with the Gaia phase spiral is approximately 0.3 to 0.9 Gyr \citep{grand2012dynamics,hunt2018transient,reid2019trigonometric}. Therefore, treating the potential as approximately time-independent on this timescale is a common approximation.
Using empirical fits to the velocity dispersions of thin-disk and thick-disk stellar populations, the velocity of 3I/ATLAS is far more probable under the thin-disk distribution. Moreover, because thin-disk stars dominate the local number density serving as a natural prior, the posterior probability that 3I/ATLAS is ejected from the thin disk is further enhanced.

Additionally, when candidate encounters are derived from the \textit{Gaia} DR3, many of the stars lack robust spectral types, age estimates, or reliable masses. This scarcity of stellar parameters further hinders the identification of host stars for 3I/ATLAS at the current stage. Machine-learning methods can be used to integrate the currently observed properties of ISOs to better infer their host stars and birth environments. In particular, by simulating both the ejection of ISOs and the physical processing they undergo during interstellar travel like cosmic-ray irradiation and space weathering, observable features such as kinematics, colors, and elemental abundances enable models to learn nonlinear correlations and to represent full probability distributions in a high-dimensional feature space. Compared with traditional orbit-integration pipelines, these models are designed to improve origin inference for ISOs.
As additional interstellar objects are discovered, a comprehensive evaluation system for determining host stars of interstellar objects  will be necessary for deepening our understanding of planetary systems beyond the Solar System.

\section*{Acknowledgments}
This work was supported by the National Key R\&D Program of China (Nos.~2024YFA1611801 and 2024YFC2207800) and by the National Natural Science Foundation of China (NSFC, Grant No.~12473066). 
It was also supported by the Shanghai Jiao Tong University 2030 Initiative and the China--Chile Joint Research Fund (CCJRF No.~2205). 
CCJRF is provided by the Chinese Academy of Sciences South America Center for Astronomy (CASSACA) and was established by the National Astronomical Observatories, Chinese Academy of Sciences (NAOC), and the Chilean Astronomy Society (SOCHIAS) to support China--Chile collaborations in astronomy. 
Additional support was provided by the Fundamental Research Funds for the Central Universities, the ''111 Project'' (No.~B20019), and the Shanghai Natural Science Foundation (Grant No.~19ZR1466800). This project was also supported by the Office of Science and Technology, Shanghai Municipal Government (grants 24DX1400100 and ZJ2023-ZD-001), and by the SJTU-Warwick Joint Seed Fund 2024/25-Round 5).
Luyao was supported by the China Scholarship Council (CSC) in collaboration with the University of Leicester. 
This work is based on data from the European Space Agency (ESA) mission \textit{Gaia} (\href{https://www.cosmos.esa.int/gaia}{\url{www.cosmos.esa.int/gaia}}), processed by the \textit{Gaia} Data Processing and Analysis Consortium (DPAC; \href{https://www.cosmos.esa.int/web/gaia/dpac/consortium}{\url{www.cosmos.esa.int/web/gaia/dpac/consortium}}).


\vspace{5mm}

% Similar to \facility{}, there is the optional \software command to allow 
% authors a place to specify which programs were used during the creation of 
% the manusscript. Authors should list each code and include either a
% citation or url to the code inside ()s when available.

% Appendix material should be preceded with a single \appendix command.
% There should be a \section command for each appendix. Mark appendix
% subsections with the same markup you use in the main body of the paper.
% Each Appendix (indicated with \section) will be lettered A, B, C, etc.
% The equation counter will reset when it encounters the \appendix
% command and will number appendix equations (A1), (A2), etc. The
% Figure and Table counter will not reset.
\clearpage

% \clearpage                       % ← ★ 关键：把所有残余浮动体全部排出
\bibliographystyle{aasjournal}   % 不写 .bst
\bibliography{ref}               % 不写 .bib；保持在 \clearpage 之后
\end{document}